# Comments on "Band gap and band parameters of InN and GaN from quasiparticle energy calculations based on exact-exchange density-functional theory" [Appl. Phys. Lett. 89, 161919 (2006)]

D. Bagayoko, L. Franklin, and G. L. Zhao
Department of Physics, Southern University and A & M College
Baton Rouge, Louisiana 70813

**Abstract**

An oversight of some previous density functional calculations of the band gaps of wurtzite and cubic InN and of wurtzite GaN by Rinke et al. [Appl. Phys. Lett. 89,161919, 2006] led to an inaccurate and misleading statement relative to limitations of density functional theory (DFT) for the description of electronic properties of these materials. These comments address this statement. In particular, they show that some local density approximation (LDA) calculations have correctly described or predicted electronic and related properties of these systems [Phys. Rev. B 60, 1563, 1999; J. Appl. Phys. 96, 4297, 2004, and 97, 123708, 2005]. *These successful calculations solved self-consistently the system of equations defining LDA, i.e., the Kohn-Sham equation and the equation giving the ground state charge density in terms of the wave functions of the occupied states.*

In their article on the "Band gap and band parameters of InN and GaN from quasiparticle energy calculations based on exact-exchange density-functional theory," Rinke et al.[1] reported results of their calculations of band gaps and related properties for wurtzite GaN (w-GaN), InN (w-InN), and cubic (i.e., zinc blende) InN (c-InN). Their computations first employed the exact-exchange optimized effective potential (OEPx) method, using density functional potentials. These OEPx calculations were reportedly performed with a plane wave and pseudopotential code. The authors subsequently employed a Green function (G) and screened Coulomb (W) approximation starting with the OEPx results. Their $G_0W_0$ calculations utilized a space-time method.[1] These quasiparticle calculations led to band gaps that agree with experiment for both wurtzite and zinc blende GaN and InN, as shown in the first table in their article.

The present comments are not a criticism of the actual calculations summarized above. Rather, they concern an oversight, that presumably led to an inaccurate and misleading statement by the Rinke and co-authors,[1] of several previous findings from density functional theory (DFT) calculations. Specifically, they stated that previous DFT calculations, in the local density approximation (LDA), predict InN to be metallic in both the wurtzite and zinc blende structures. This assertion is inaccurate in light of previous DFT and LDA results for w-GaN,[2] c-InN,[3] and w-InN.[4] Table 1 below shows these previous, calculated DFT-LDA results that are in excellent agreement with experiment,[5-8] for these systems. For c-InN, the 2004 theoretical predictions[3] from *ab-initio*, self-consistent, LDA calculations have recently been confirmed by experiment,[5] for both the lattice constant and the band gap.

The above inaccurate statement, in light of the content of Table 1, is also misleading. The distinction of the noted DFT and LDA calculations as compared to others stems from their implementation of the Bagayoko, Zhao, and Williams (BZW) method.[2-4,9] This rigorous method systematically searches for the optimal basis set that is *complete vis-à-vis* the description of the ground state of the material under study. *Doing so amounts to solving self-consistently both the Kohn-Sham equation and the equation giving the ground state charge density in terms of the wave functions of the occupied states.* It entails no ad hoc corrections to the band gap and no additional parameters other than the ones inherent to the applicable LDA potential.

Table 1. Band gaps (Eg), in eV, of w-GaN, w-InN, and c-InN. In light of the Burstein-Moss effect in w-InN, we note that the experimental results of References 6, 7, and 8 are respectively for carrier densities of 5.48 $10^{18}$, 1.2 $10^{19}$, and 5 $10^{19}$ cm$^{-3}$. The lattice constants below are in Angstroms.

| Systems and Computational Method | $a_0$ (Å) | $c_0$ (Å) | u | Eg (eV) |
| --- | --- | --- | --- | --- |
| w-GaN (OEPx-$G_0W_0$)[a] | 3.181 | 5.166 | 0.377 | 3.32[a] |
| (LDA-BZW)[b] | 3.160 | 5.125 | 0.377 | 3.2[b] |
| w-InN (OEPx-$G_0W_0$)[a] | 3.533 | 5.693 | 0.379 | 0.72[a] |
| (LDA-BZW)[c] | 3.544 | 5.718 | 0.379 | 0.88[c] |
| (GGA-BZW)[c] | 3.544 | 5.718 | 0.379 | 0.81[c] |
| Experiments | | | | 0.7-0.8[d], 0.883[e], 0.89[f] |
| c-InN (OEPx-$G_0W_0$)[a] | 4.98 | | | 0.53[a] |
| (LDA-BZW)[g] – predictions | 5.017[g] | | | 0.65[g] |
| (LDA-BZW)[g] | 4.98 | | | 0.74[g] |
| Experiment[h] - confirmation | 5.01±0.01[h] | | | 0.61[h] |

[a] Reference 1   [b] Reference 2   [c] Reference 4   [d] Reference 6   [e] Reference 7   [f] Reference 8

[g] Reference 3   [h] Reference 5

The BZW method avoids *a basis set and variational* effect[2-4,9-11] that is inherent to self-consistent calculations employing a basis set in a variational approach of the Rayleigh-Ritz type. These self-consistent calculations need not employ DFT (or LDA) potentials for this assertion to hold. In addition to the excellent agreement between experiment and the LDA-BZW results for the electronic structures and band gaps of numerous semiconductors,[2-4,9-11] we should note the overall correct description of low-lying unoccupied energy bands.

This result is apparent from the agreement, with experiment, of LDA-BZW calculated optical properties of selected materials.[9-10] The method reproduced the measured dielectric functions of barium titanate (BaTiO$_3$)[9] and of w-InN.[10] For the latter, the agreement with experiment is up to photon energies of 6 and 5.5 eV for the imaginary and real parts of the dielectric function, respectively. An added feature of the LDA-BZW results stems from their agreement with experiment for the electron effective masses of the systems studied.[2-4,9,11] These effective masses are measures of the correctness of the curvature (or shape) of the lowest lying conduction band.

In summary, these comments are intended to correct an oversight of several, previous LDA-BZW results that agree very well with experiment. This oversight is believed to have led to an inaccurate statement relative to density functional theory description of w-InN and c-InN. *Our computational method simply follows the explicit statement by Kohn and Sham[12] and by Kohn[13] relative to the need to solve self-consistently the system of equations describing LDA.*

**References**

[1] P. Rinke, M. Scheffler, A. Qteish, M. Winkelnkemper, D. Bimpberg, and J. Neugebauer, Appl. Phys. Lett. 89, 161919 (2006).
[2] G. L. Zhao, D. Bagayoko, and T. D. Williams, Phys. Rev. B **60**, 1563 (1999).
[3] D. Bagayoko, L. Franklin, and G. L. Zhao, J. Appl. Phys **96**, 4297 (2004).
[4] D. Bagayoko, and L. Franklin, J. Appl. Phys. **97**, 123708 (2005).
[5] J. Schörmann, D. J. As, K. Lischka, P. Schley, R. Goldhahn, S. F. Li, W. Löffler, M. Hetterich, and H. Kalt, Appl. Phys. Lett. **89**, 261903, 2006.
[6] J. Wu et al., Appl. Phys. Lett. 80, 3967 (2002).
[7] J. Wu et al., J. Appl. Phys. **94**, 4457 (2003).
[8] T. Inishima, V. V. Mamutin, V. A. Vekshin, S. V. Ivanov, T. Sakon, M. Motokawa, and S. Ohoya, J. Cryst. Growth **227-228**, 481 (2002).
[9] D. Bagayoko, G. L. Zhao, J. D. Fan, and J. T. Wang, J. Phys. Cond. Matt. Vol. 10, No. 25, 5645 (June, 1998).
[10] H. Jin, G. L. Zhao, and D. Bagayoko, J. Appl. Phys. **101**, 033123 (2007).
[11] H. Jin, G. L. Zhao, and D. Bagayoko, Phys. Rev. B **73**, 245214 (2006).
[12] W. Kohn and L. J. Sham, Phys. Rev. 140, A1133 (1965).
[13] W. Kohn, Reviews of Modern Physics, Vol. 71, No. 5, 1253 (1999). Nobel Lecture of Professor Walter Kohn; also available at http://nobelprize.org/nobel_prizes/chemistry/laureates/1998/kohn-lecture.pdf